\documentclass[preprint]{revtex4}

\usepackage{graphicx}
\newcommand{\eg}{{\it e.g.}}

\begin{document}
\bibliographystyle{revtex}

\preprint{IIT-HEP-01/6}

\title{The BTeV Vertex Trigger System\footnote{Presented at the {\sl Snowmass Summer Study on the Future of Particle Physics}, Snowmass, Colorado, June 30--July 21, 2001.}
}



\author{Daniel M. Kaplan}
\email[]{kaplan@fnal.gov}
\affiliation{Illinois Institute of Technology, Chicago, IL 60616}

\author{for the BTeV Collaboration}
\noaffiliation


\begin{abstract}
The BTeV trigger system will be very ambitious, reconstructing almost all tracks in real time at 15 million 2-TeV ${\bar p} p$ interactions per second. The goal is substantially-increased sensitivity to rare effects in heavy-quark physics over that of existing experiments. The evolution of our approach is briefly summarized.
\end{abstract}

\maketitle

\section{Introduction}

BTeV is an approved Fermilab experiment that will use the Tevatron Collider to study heavy-quark physics. BTeV data-taking is planned to commence in Tevatron Run IIB (starting about 2006). To achieve the highest possible sensitivity to mixing, {\em CP} violation, and rare decays, a very ambitious trigger system is envisioned that will examine {\em every} beam crossing for evidence of secondary vertices. 

The trigger is conceived in three levels.
The Level 1 trigger (the lowest trigger level) will employ custom pattern-recognition hardware to perform fast trackfinding on every beam crossing, using information from the silicon-pixel vertex detector~\cite{BTeV-proposal,FPIX2}. Events for which the pixel detectors show evidence of secondary vertices will be passed on to Level 2, which will use information from additional tracking detectors to improve the vertex resolution and refine the Level 1 decision. Events passing Level 2 will undergo a more complete event reconstruction at Level 3, with information available from all detectors. (In addition we are building a high-$p_t$-muon trigger, which we do not discuss further here; similar hardware will be used to implement both triggers.)
 
\section{Vertex trigger strategy}
\begin{figure}
\centerline{\rotatebox{90}{\scalebox{0.4}{\includegraphics{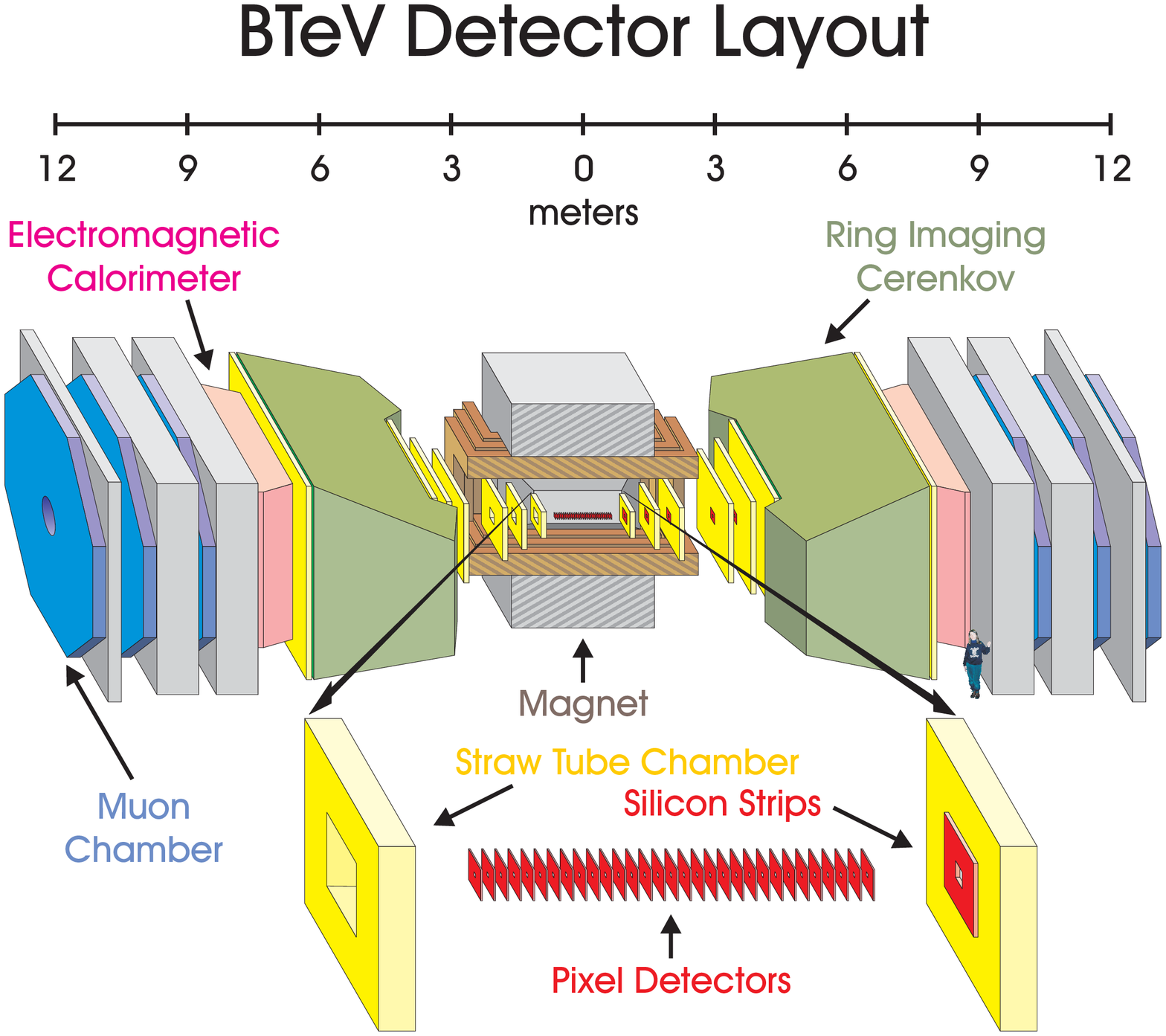}}}}
\caption{Layout of BTeV spectrometer; note that the pixel vertex detector is located within the dipole spectrometer magnet centered on the interaction region.}
\label{fig:BTeV}
\end{figure}

\begin{figure}
\centerline{\scalebox{0.4}{\includegraphics{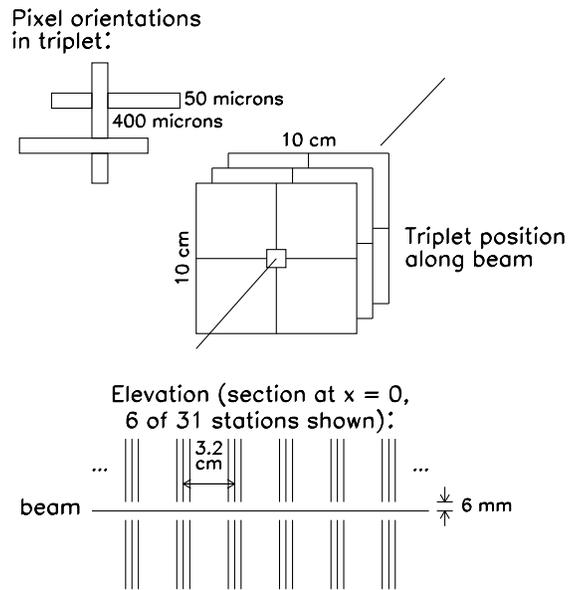}}}
\caption{Details of an early BTeV pixel-detector layout: ``triplet scheme" proposed in the BTeV ``Preliminary TDR"~\protect\cite{BTeV-pTDR}.}
\label{fig:pixels}
\end{figure}

BTeV is designed for a 7.6-MHz beam crossing rate at an average of 2 interactions per crossing.
In reconstructing tracks in real time from all crossings, the key problem is the combinatoric processing time that typically limits the speed of trackfinding algorithms. Since the initial conception of BTeV~\cite{BTeV-EoI}, our thinking has evolved on the best approach for meeting this challenge.
A key part of the solution is to optimize the design of the vertex detector not only for good vertex resolution, but also for fast pattern recognition. For this reason we are designing thin silicon pixel detectors with fast, data-driven readout~\cite{FPIX2} and locating them within the spectrometer magnet, so that momentum information is available already at the lowest trigger level (see Fig.~\ref{fig:BTeV}). Fast readout  with high hit-rate capability requires rectangular pixels (Fig.~\ref{fig:pixels}), to maintain excellent measurement resolution in at least one coordinate while providing sufficient room behind each pixel for the complex pixel signal-processing and readout circuitry. 

Our initial conception of the trigger algorithm~\cite{Kaplan-B99}, based on work by the Penn group~\cite{Husby}, required that each pixel station by itself determine a short track segment or ``mini-vector." The mini-vector should have sufficient resolution to select  only a small search region in the next station, with mean occupancy well below 1 hit per crossing. To accomplish this, we 1)~provided each station with three pixel planes and 2)~divided each plane logically into 32 pie-shaped sectors (``$\phi$-slices").   Monte
Carlo simulation gave a mean hit occupancy
of 0.23/interaction/$\phi$-slice. By providing in each station two ``precision-$y$" planes and one ``precision-$x$" plane, we could determine, using the hits in a single station, a mini-vector 
accurate to $\approx\!\pm1\,$mr in $y$ and $\pm50\,$mr in
$x$ that would thus project into a very small fraction ($\approx$
$2\times10^{-5}$) of the area of the next station, within which the occupancy due to other tracks or random hits would be negligibly small.

This vertex-detector strategy led to a high-performance, cost-effective trigger implementation~\cite{BTeV-pTDR,Kaplan-B99}. However, it required an extra pixel plane per station compared to the minimum needed for off-line event reconstruction. In view of the total thickness of the triplet-scheme vertex detector (approaching 1 radiation length), as well as its cost, a 2-plane-per-station solution was clearly desirable, and in 1999 we embarked on a  two-year exploration of alternative trigger approaches that would be compatible with 2 planes per station.
Two-plane stations required giving up the mini-vector approach; the key trigger
problem then becomes the trackfinding combinatorics. 

We have studied two alternative approaches compatible with two-plane stations.  At 2 interactions/crossing, a pixel-plane quadrant has 8 hits  on average; the probability of $>$16 hits is $<$1\%. The first alternative algorithm we considered used a massively-parallel array of line-segment finders (based \eg\ on Xilinx gate arrays) that could process the resulting large number of combinations in $<$132\,ns/crossing. The linefinders would  look for hit triplets in three successive stations in each view, using linear interpolation since track curvature is negligible over the several-cm 3-station separation. This approach might be viewed as replacing detector slicing in $\phi$ with slicing in $z$: a separate linefinder would be provided for each station, all running in parallel, and a subsequent step would link the track segments together into tracks, find the primary vertices, and identify any tracks that miss the primary.

\begin{figure}
\centerline{\scalebox{0.75}{\includegraphics{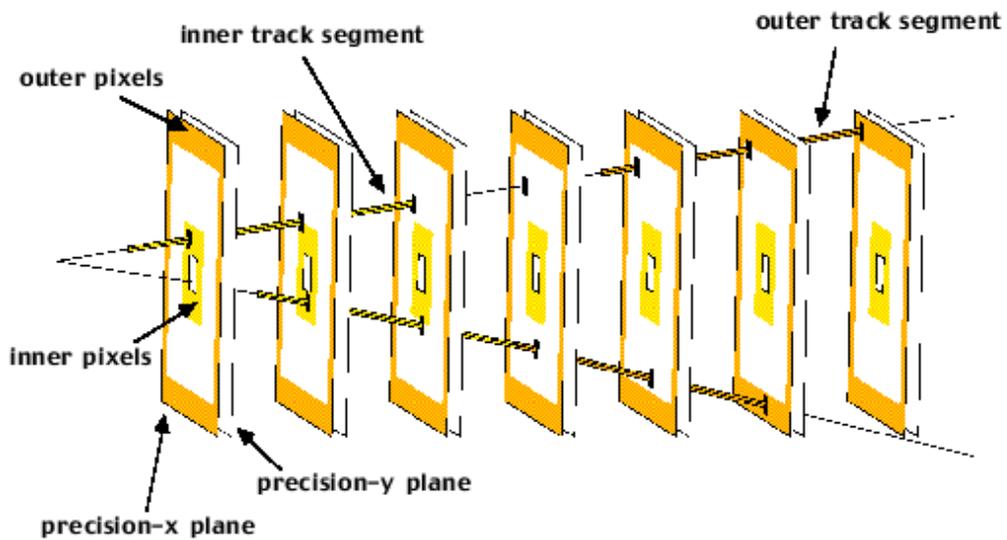}}}
\caption{Illustration of some features of the doublet trackfinding scheme.}
\label{fig:inner-outer}
\end{figure}

The two-plane algorithm just described would find each track many times, since it looks in all stations in parallel. We have also explored alternative approaches that reduce the total processing power needed by finding each track only once near the point of optimal vertex resolution. Such an algorithm~\cite{Erik} is the one we now propose to use~\cite{BTeV-proposal}. In this approach, pixel chips closest to the beam are classified as ``inner" and those farthest from the beam (near the periphery of each pixel plane) ``outer,"  and only those track segments are sought that include inner pixels or outer pixels (see Fig.~\ref{fig:inner-outer}). A track segment including an inner pixel (inner segment) is required to extrapolate into the beam hole at the $z$ of the previous station, and one including an outer pixel (outer segment) is required to extrapolate beyond the vertex detector at the $z$ of the next station. To suppress ghost tracks, each inner segment must be confirmed by a corresponding outer segment. Each track is thus found once, near its points of entry into and exit from the vertex detector, and finding the remaining hits of the track, which do not provide essential information at this stage (although they {\em can} be used to improve the precision of the vertex fit), is postponed until Level 2. This algorithm can be implemented in an array of about 500 large FPGAs. The algorithm has been simulated and finds $>$90\% of tracks while keeping the number of ghost tracks to an acceptable level.

Following the FPGA trackfinding stage just described, tracks are grouped into primary (interaction) vertices in an array of processor farms, comprising about 2500 DSPs in our current baseline design. Tracks with large impact parameters with respect to these vertices are evidence of  heavy-quark decays. To reject light-quark background, events are required to contain $n$ tracks missing a primary vertex by at least $m$ standard deviations.  (To avoid excessive false triggers from multi-interaction crossings, only impact parameters $<$2\,mm are accepted.) A requirement such as $n=2, m=6$ rejects 99\% of light-quark events while maintaining efficiency well above 50\% for most $B$ decays of interest. The trigger system will have sufficient flexibility to allow a variety of such $n,m$ requirements to be used simultaneously, each with its own prescale factor.

The Level 2 and 3 triggers can be implemented in farms of commercial general-purpose processing nodes; we anticipate about 2500 such nodes in Levels 2 and 3. The goal of Levels 2 and 3 is to provide an additional factor of $\approx$100 in data rate while degrading efficiency for events of interest only minimally ($<$10\%). To accomplish this, most remaining background events will be rejected, with much of the raw data in the remaining events summarized to reduce event size.

\section{Trigger-DAQ Integration}

The data rates into and out of the trigger levels are large (Table~\ref{tab:rates}) and require close integration between the trigger and data-acquisition systems. These rates are carried by about 2500 1\,GByte/s data links into and out of about  1\,TByte of buffer memory. Detectors all read out directly into the Level 1 buffer, which passes the pixel data to the Level 1 trigger processors. For events failing Level 1, all data are flushed, while for those events that pass, additional tracking data are then provided from the Level 1 buffers to the Level 2/3 farms. As events move into the later stages of Level 3, data from additional detectors are provided to allow the decision to be further refined. Finally, events satisfying all trigger levels are compressed and written to an archival medium.

\begin{table}
\caption{Estimated data rates at various points in the BTeV Trigger/DAQ system.}
\label{tab:rates}
\centering
\begin{tabular}{lc}
\hline
Where & How much \\
\hline
From detector & 1.5\,TB/s \\
Into Level 1 trigger & $\approx$0.1\,TB/s \\
Into Level 2/3 trigger & $\approx$25\,GB/s \\
To archival medium & $\approx$200\,MB/s \\
\hline
\end{tabular}
\end{table}

\section{Conclusions}
To achieve unprecedented sensitivity to rare effects in charm and beauty decay, the BTeV collaboration has proposed to build the most sophisticated trigger system yet attempted in a high-energy-physics experiment. While mutiple reviews have revealed no show-stoppers, we realize that the task is challenging. Even downloading, control, monitoring, and calibration will require the development of new techniques~\cite{CHEP}.


\end{document}